\documentstyle[11pt,newpasp,twoside,psfig]{article}
\index{summary}
\index{instructions}
\index{template}

\def\axp{\hbox{1E 1841$-$045}}

\newcommand\xte{{\it RXTE}}

\newcommand\ginga{{\it Ginga}}

\newcommand\sax{{\it Beppo}SAX}

\newcommand\asca{{\it ASCA}}

\def\sp{\hskip 1.5pt}

\def\kes73{\hbox{Kes\sp73}}

\def\edcomment#1{\iffalse\marginpar{\raggedright\sl#1\/}\else\relax\fi}
\reversemarginpar

\begin{document}
\title{\hbox{15 Years Going Steady -- Timing the Magnetar~1E~1841-045}}

\author{E. V. Gotthelf }
\affil{Columbia Astrophysics Laboratory, 550 West 120th St, New York, NY 10027, USA}
\author{F. P. Gavriil, V. M. Kaspi} 
\affil{McGill University, 3600 University St, Montreal, QC, H3A~2T8, CA}
\author{G. Vasisht}
\affil{JPL/NASA, 4800 Oak Grove Drive, Pasadena, CA, 91109, USA}
\author{Deepto Chakrabarty}
\affil{MIT, 70 Vassar Street, Cambridge, MA 02139, USA}

\begin{abstract}
We report on a long-term monitoring campaign of \axp, the 12-s
anomalous X-ray pulsar and magnetar candidate at the center of the
supernova remnant \kes73.  We have obtained approximately monthly
observations of the pulsar with the {\it Rossi X-ray Timing Explorer}
(\xte) spanning over two years, during which time \axp\ is found to be
rotating with sufficient stability to derive a phase-connected timing
solution. A linear ephemeris is consistent with observations of the
pulse period made over the last 15~yrs with the \ginga, \asca, \xte,
\& \sax\ observatories.  Phase residuals suggest the presence of
``timing noise'', as is typically observed from young radio pulsars.
These results confirm a rapid, constant spin-down for the pulsar,
which continues to maintain a steady flux; this is inconsistent with
most accretion scenarios.
\end{abstract}

\section{Introduction}
1E~1841$-$045 is perhaps the best candidate for a magnetar -- an
isolated neutron star (NS) with an extreme magnetic field whose energy
loss is dominated by magnetic field decay (Vasisht \& Gotthelf 1997;
see Duncan \& Thompson 1992 for theory of magnetars).  This
interpretation is based primarily on initial studies which showed that
the pulsar is slowing down rapidly ($\dot P = 4.1 \times 10^{-11}$
s/s), with an inferred equivalent magnetic dipole field of $B_p \sim
7.0 \times 10^{14}$ G, nearly twenty times the quantum critical field,
while producing steady emission at a rate far in excess of its
rotational kinetic energy loss (Vasisht \& Gotthelf 1997). Further
evidence for an isolated system is the absence of Doppler shifts,
detectable companion or accretion disk, and a spectrum which differs
greatly from those of known accretion-powered binary systems.

Here we report on a two-year monitoring campaign of \axp\ made with
\xte. We have obtained a spin ephemeris derived from a phase-connected
timing solution which is found to be consistent with nearly two
decades of observations of the pulsar. We are able to characterize
rotational stability of \axp, including a search for timing noise,
pulsed flux variability, and frequency glitches. These results provide
important observational constraints on both the magnetar and any
accretion model for \axp.

\section{Observations and Results}

We acquired a total of 45 observations of \axp, spaced nearly
uniformly, during the interval spanning 1999 February 15 through 2001
April 2, with a typical observation lasting $\sim$7~ks. These
observations were made with the Proportional Counter Array (PCA;
Jahoda et al. 1996) aboard \xte.  The PCA consists of 5 collimated
proportional counter units (PCUs) with a total effective area of $\sim
6500$~cm$^{2}$ in the energy range $2-60$~keV and a resolution of
$\sim 18\%$ at 6 keV over its $\sim1^{\circ}$ field of view (FWHM).
 
Data were collected in {\tt GoodXenonwithPropane} data mode with the
photon arrival times recorded at 1~$\mu$s resolution and the energy
binned into 256 spectral channels.  To maximize the signal-to-noise
ratio of the pulsar given the spectral properties of the source and
background we restricted the energy range to 2.1--5.4~keV and used
events from the top Xenon layer only.  All arrival times were
corrected to the solar system barycenter.

To determine the average pulse time-of-arrival (TOA) for each
observation we folding the photon arrival times into 32 bins modulo
the test frequency. This frequency was determined initially from a
Fast Fourier transform and later from an approximate timing
ephemeris. The resulting pulse profiles were then cross-correlated
with a high signal-to-noise template and the measured TOAs fitted
to a polynomial using the {\tt TEMPO} timing software package.  The
uncertainty on the best-fit model parameters, the pulse frequency
$\nu$ and its time derivatives, were small enough to allow prediction
of the phase of the next observation to within $\sim 0.2$,
given the spacing of our monitoring observations. This is
only practical for a very stable rotator.

Figure 1 presents the results of our fit. Complete modeling of the
pulse arrival times required four time derivatives of $\nu$ to produce
residuals from this fit whose RMS variability are comparable with the
TOA arrival times uncertainties. These residuals, defined as the
difference between the observed and model-predicted pulse arrival
times, have RMS deviations of only 3\% of the pulse period. We find
that \axp\ has maintained phase coherence over the $2$~year duration
of our observations with no indication of any glitch activity, i.e.,
no sudden changes in the pulse period.  The data are well modeled for
all but three observations which took place near the start of our
monitoring campaign. These $\sim 5 \sigma$ points, however, deviate by
only a small fraction of the pulse period.  For all cases, we
estimated the uncertainties on the pulse arrival times by means of
Monte-Carlo simulations of the pulse profiles.

In Figure 2, we compare our derived ephemeris with pulse frequency
measurements spanning 15~yrs obtained with the \ginga, \asca, \xte, \&
\sax\ observatories as reported in Gotthelf et al. (1999). Despite the
need for four frequency derivatives to fully model the pulse arrival
times, the linear ephemeris ``predicts'' the historic data consistent
with their measured errors; in contrast, when higher order derivatives
are included in the ephemeris, large departures from the earlier data
points result. We conclude that these higher order terms ($>
\dot{\nu}$) are likely random wandering of the timing residuals about
an otherwise very stable, constant spin-down. Timing residuals of this
nature are often measured for radio pulsars and are generally referred
to as radio ``timing noise'' (see, e.g., Cordes \& Helfand 1980).  The
magnitude of second derivative gives the strength of the timing noise
observed from \axp.

\begin{figure}[t]
\small
      \begin{minipage}[t]{0.48\linewidth}
\psfig{file=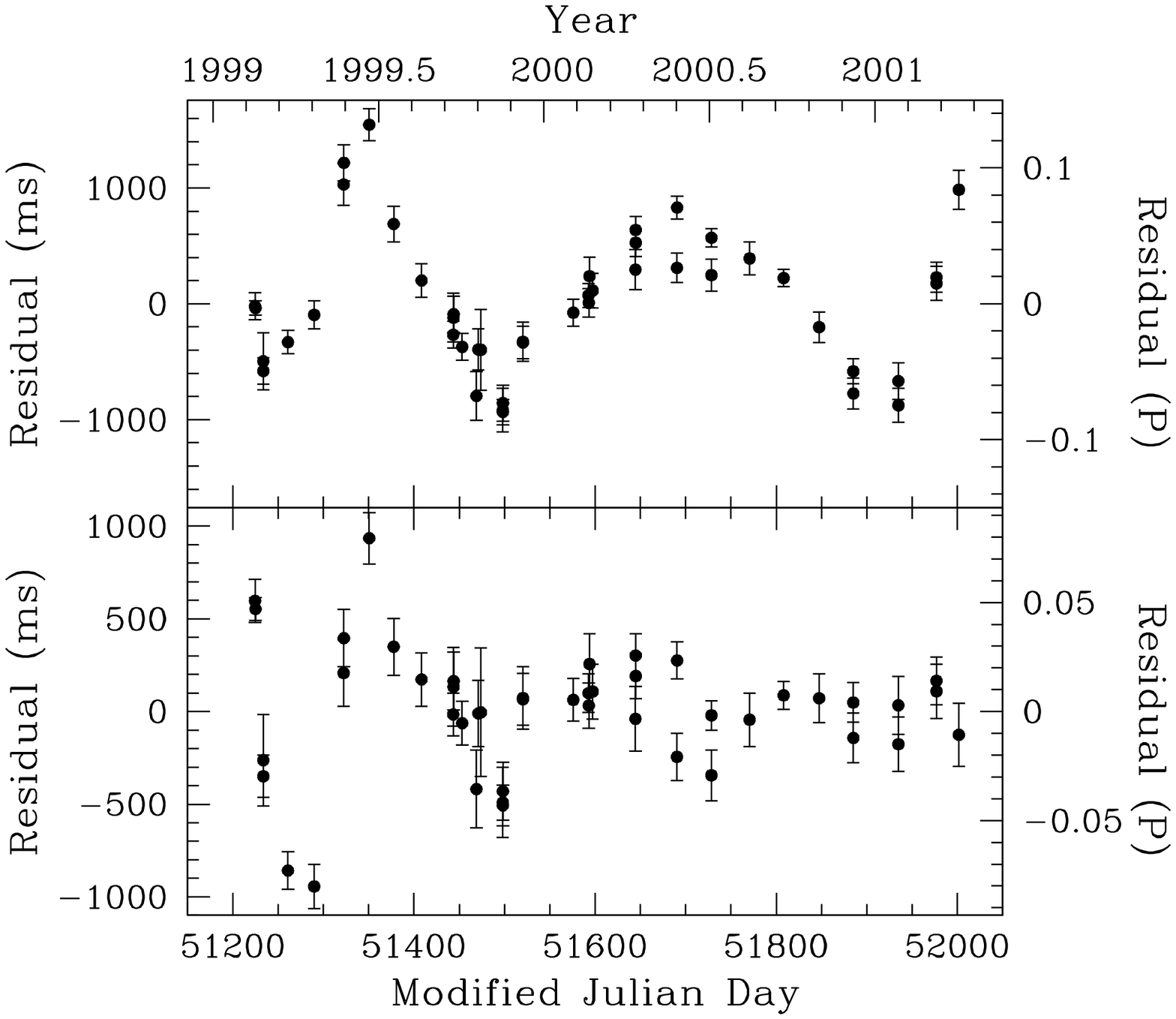,width=2.6in,angle=0.0} 
{
{\bf Figure 1.} Arrival time residuals for the 12~s pulsar \axp\ in
Kes~73 versus epoch. (Top panel) Residuals after removing the leading
3 frequency derivative terms from the phase-connected \xte\ ephemeris.
Evidently there is much ``timing noise'' characteristic of young
radio pulsars. (Bottom panel) The arrival time residuals using an 
ephemeris with four frequency derivatives. The RMS
residuals from the final fit is $\pm3\%$ of the pulse period. Three
data points are found to differ from the mean by $\sim 5\sigma$.
}
      \end{minipage}\hfill
      \begin{minipage}[t]{0.48\linewidth}
        \psfig{file=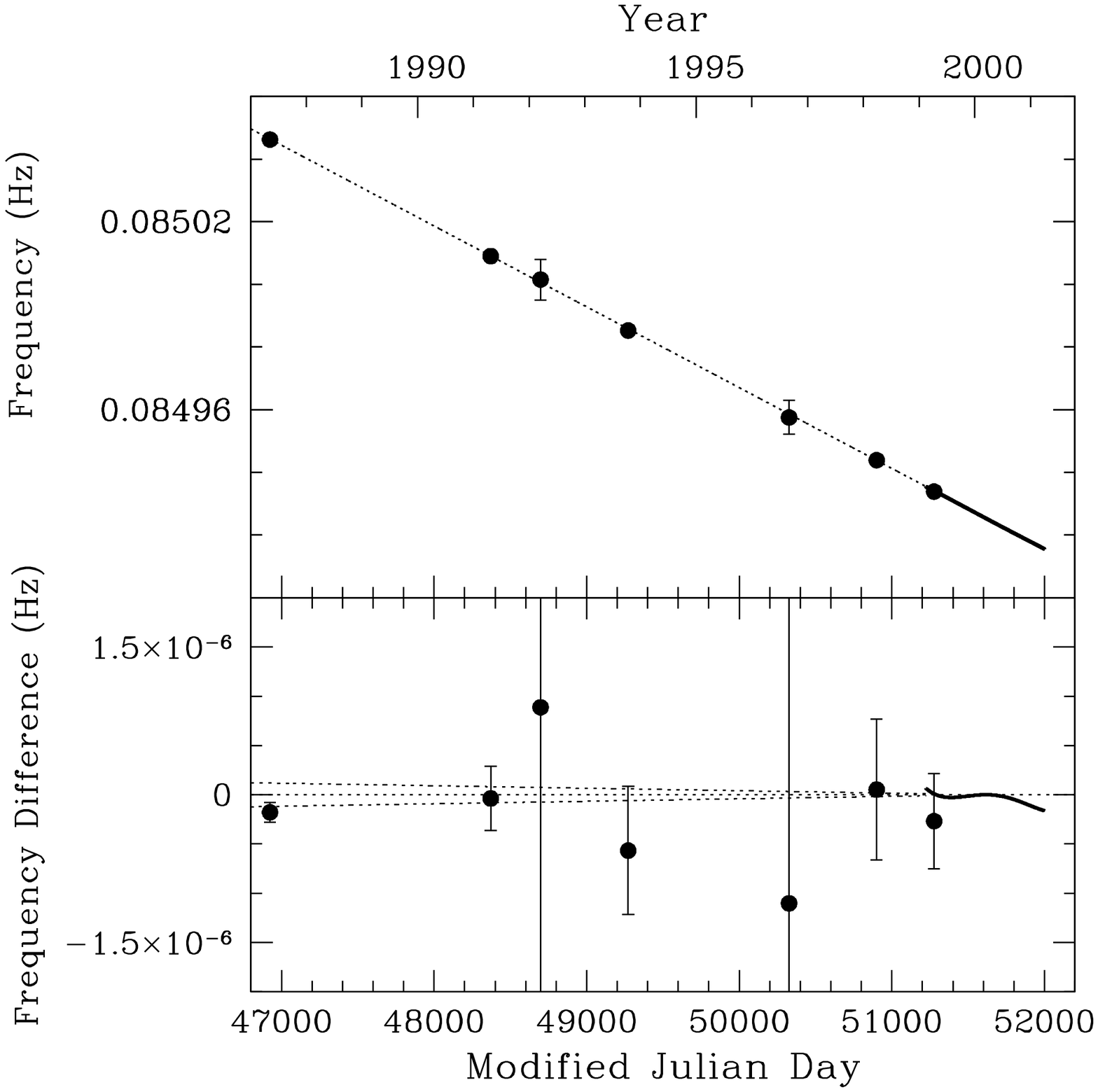,width=2.6in,angle=0.0}
{
{\bf Figure 2.} The spin-down history of \axp\ and its residuals from a
{\it linear} ephemeris (thick solid line). The filled circles show the
previously observed spin frequencies (from Gotthelf et al. 1999).  The
linear terms of the phase-connected ephemeris, when extrapolated over
10 years (dotted line), is found to be consistent with the
previous measurements, despite its appreciable timing noise.
The uncertainty in the
extrapolation is comparable the width of the dotted line.  }
\end{minipage}
\end{figure}

\section{Discussion}

Observationally \axp\ is characterized as an Anomalous X-ray Pulsar
(AXP; see Mereghetti 2001 for a review). This small group ($\sim 6$
members) of seemingly isolated pulsars are observed exclusively in the
X-ray energy band\footnote{There is recent evidence for IR emission
from two AXPs, see Hulleman et al 2000, 2001.}, are underluminous for
an accreting X-ray binary system, and are, except for the
characteristic burst activity, otherwise similar observationally to
the SGRs.  This is the fourth AXP, after 1E~2259+586, 1E~0142+615, and
RXS J1708$-$4009, for which long-term phase-connected timing is
possible, characterizing these objects as rotators of great stability
(see Gavriil \& Kaspi 2001).  In contrast, the AXP 1E~1048.1$-$5937 is
much less stable, as a phase-connected timing solution cannot be
maintained for more than a few months (Kaspi et al. 2001).  The
measured strength of the torque noise for \axp\ is comparable to that
found for 1E 1048.1$-$5937, during its intervals of relative
stability, but substantially weaker than the torque noise measured for
most accreting pulsars (Bildsten et al. 1997). Still, it is a factor
of two quieter than even the most exceptionally quiet accreting
pulsars (e.g., 4U 1626$-$67, Chakrabarty et al. 1997).

The leading theory for the nature of AXPs is the magnetar model as
first proposed by Thompson \& Duncan (1996).  In this model, in the
absence of soft-gamma-repeater-like outbursts, one expects generally
smooth spin-down.  The lack of deviations from a simple spin-down
model found for \axp\ is consistent with the magnetar model.  An
alternative model proposed for AXPs is that they are accreting from a
disk of material that formed shortly after the supernova explosion
that gave birth to the neutron star.  Severe constraints have been
placed on the plausibility of this scenario by optical/IR observations
of AXPs (Hulleman et al. 2000).  Both the constant, long-term
spin-down of \axp, as well as the fact that phase-connected timing is
possible for this source, provide evidence against accretion
scenarios, as spin-up episodes, as well as considerably more torque
noise, are in general to be expected.
   
\acknowledgments

{This research is supported by NASA LTSA grants NAG~5-22250 and NAG~5-8063, NSERC Rgpin 228738$-$00, and Sloan Fellowship. V.M.K. is a Canada Research Chair.}


\begin{references}
 
\reference Bildsten, L. et al. 1997, ApJS, 113, 367

\reference Chakrabarty, D. et al. 1997, ApJ, 474, 414

\reference Cordes, J. M. \& Helfand, D. J. 1980, ApJ, 239, 640

\reference Duncan, R. C. \& Thompson, C. 1992, ApJ, 392, 9

\reference Gotthelf, E. V., Vasisht, G. 1997, ApJ, 486, L133

\reference Gotthelf, E. V., Vasisht, G., Dotani, T. 1999, ApJ, 522, L49

\reference Hulleman, F., van Kerkwijk, M. H., Kulkarni, S. R. 2000, Nature, 408, 689

\reference Hulleman, F. et al. 2001, in press

\reference Jahoda, K., Swank, J.~H., Giles, A.~B., Stark, M.~J., Strohmayer, T., Zhang, W., \& Morgan, E.~H. 1996, Proc. SPIE, 2808, 59

\reference Kaspi, V.~M., Chakrabarty, D., \& Steinberger, J. 1999, ApJ, 525, L33

\reference Kaspi, V. M., Gavriil, F. P., Chakrabarty, D., Lackey, J. R., Muno, M. P. 2001, ApJ, 558, 253

\reference Gavriil, F. P., \&  Kaspi, V. M.  2001, ApJ, in press

\reference Mereghetti, S. 2001, NATO ASI Ser., Dordrecht Kluwer, in press

\reference Thompson, C. \& Duncan, R. C. 1996, ApJ, 473, 322

\reference Vasisht, G. \& Gotthelf, E. V. 1997, ApJ, 486, L129

\end{references}
\end{document}